\documentclass[twoside,12pt]{article}
\usepackage{epsfig}

\newcommand{\be}{\begin{equation}}
\newcommand{\ee}{\end{equation}}
\newcommand{\bea}{\begin{eqnarray}}
\newcommand{\eea}{\end{eqnarray}}

\begin{document}

\title{ \vspace{1cm} Direct Reactions
with Exotic Nuclei, Nuclear Structure and Astrophysics
}
\author{G.Baur,$^1$ S.Typel$^2$
\\
$^1$Institut f\"ur Kernphysik, Forschungszentrum J\"ulich\\
D-52425 J\"ulich, Germany\\
$^2$ Gesellschaft f\"ur Schwerionenforschung mbH(GSI), Planckstrasse 1\\
D-64291 Darmstadt, Germany\\
}
\maketitle
\begin{abstract} 
Intermediate energy Coulomb excitation and dissociation is
a useful tool for nuclear structure and astrophysics studies.
Low-lying strength in nuclei far from stability was
discovered by this method. The effective range theory
for low-lying strength in one-neutron halo
nuclei is summarized and extended to two-neutron halo nuclei.
This is of special interest in view of recent rather accurate
experimental results on the low-lying electric dipole strength
in $^{11}$Li.
Another indirect approach to nuclear astrophysics is the 
Trojan horse method.
It is pointed out that it is 
a suitable tool to investigate subthreshold resonances.
\end{abstract}


\section{Introduction and Overview}


With the exotic beam facilities all over 
the world - and more are to come - direct reaction theories 
are experiencing a renaissance. 
These direct reaction theories are essential to interprete
indirect methods used in nuclear astrophysics. 
We discuss Coulomb dissociation \cite{bbr} and the Trojan horse method,
for a recent review see \cite{enam}.
Coulomb dissociation was reviewed in 2003
\cite{ppnp}, a comprehensive
paper on the Trojan horse method appeared in 
\cite{tyba02}.  
A more recent (mini)review of indirect methods
can be found in \cite{mumbai}.
Here we report on the effective range theory of 
low-lying electromagnetic strength in
one-nucleon halo nuclei \cite{tybaprl,tyba04},
and its extension to two-neutron halo nuclei.
This will be of special interest in view of recent rather accurate
experimental results on the low-lying electric dipole strength
in $^{11}$Li \cite{naka}.

The Trojan-Horse Method \cite{tyba02} is a particular case
of transfer reactions to the continuum under quasi-free scattering
conditions.
Special attention is paid to the 
transition from reactions to bound and unbound states
and the role of subthreshold resonances \cite{sanser}.
Since the binding energies of nuclei close to the drip
line tend to be small, this is expected to be 
an important general feature in  
exotic nuclei. 

\section{Effective Range Theory of Halo Nuclei}
\subsection{Single particle halo nuclei}
At low energies the effect of the nuclear potential is 
conveniently described by the effective-range expansion
\cite{Bet49}.
An effective-range approach for the electromagnetic
strength distribution in neutron halo 
nuclei was introduced in \cite{tybaprl,tyba04} and
applied to the single neutron halo nucleus ${}^{11}$Be.
Recently, the same method was applied to the description of
electromagnetic dipole strength in ${}^{23}$O
\cite{Noc04}.
A systematic study sheds 
light on the sensitivity of the electromagnetic 
strength distribution to the interaction in the continuum.
We expose the dependence on the binding energy of the nucleon
and on the angular
momentum quantum numbers. Our approach extends the familiar textbook
case of the deuteron,
that can be considered as the
prime example of a halo nucleus, to arbitrary nucleon+core systems,
see also \cite{kala}.
We also investigate in detail the square-well potential model.
It has  great merits: it can be solved analytically,
it shows the main characteristic features
and it leads to rather simple and transparent formulae.
As far as we know, some of these formulae have not been published before.
These explicit results can be compared to our general
theory for low energies (effective-range approach) and also 
to more realistic Woods-Saxon models. Due to shape independence,
the results of these various approaches will not differ for
low energies. It will be interesting to delineate
the range of validity of the simple models.  

Our effective-range approach is closely related to effective field theories
that are nowadays used for the description of 
the nucleon-nucleon system and halo nuclei
\cite{Ber02}. 
The characteristic low-energy
parameters are linked to QCD in systematic expansions.
Similar methods are also used in 
the study of exotic atoms ($\pi^-A$, $\pi^+\pi^-$, $\pi^-p$, \dots) 
in terms of effective-range parameters.
The close relation of effective field theory to the effective-range
approach for hadronic atoms was discussed in Ref.\ \cite{Hol99}.

In the  ANC method \cite{muka}
it is assumed that the radial continuum wave 
function $g_{l_f}$is a pure Coulomb wave
function (a regular spherical Bessel function for neutrons)
and that the dominant contribution to the radial integral 
\begin{equation}
I_r
=\int_0^{\infty}f_{l_i}(r) r g_{l_f}(r) dr
\end{equation}
comes from the region $r>R$ where R is the range of the nuclear interaction.
In this case
the S-factor is entirely determined by
the asymptotic normalization constant (ANC) $C_{l_i}$
of the bound state wave function
\begin{equation}
f_{l_i}(r) \sim C_{l_i} e^{-qr}
\end{equation}
where $l_i$ is the orbital angular momentum.
The ANC is determined by a suitable transfer reaction.
In some cases this is a good approximation.
In general there is an influence from the final state interaction
on the final continuum wave function. In the outside region
$(r>R$) it is given by
\begin{equation}
g_{l_f}=F_{l_f}\cos \delta_{l_f} + G_{l_f}\sin \delta_{l_f} 
\end{equation}
where the regular and irregular Coulomb wave functions are denoted
by $F_{l_f}$ and $G_{l_f}$ respectively.
 For the low energies
relevant now, the phase shift $\delta_{l_f}$ is given by the 
low energy parameters of the effective range expansion.
Usually this is one number, the scattering length.

\begin{figure}
\includegraphics[height=.3\textheight]{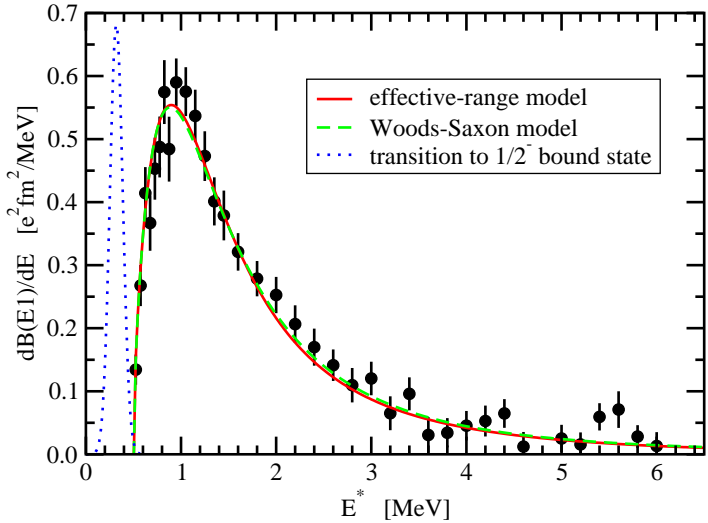}
\caption{\label{fig:1} The
reduced electric dipole transition probability
is shown as a function
of the excitation energy $E^{\ast}=E+S_{n}$.
The theoretically calculated dipole strength is
compared to experimental data extracted from the
Coulomb dissociation of ${}^{11}$Be \cite{palit}.}
\end{figure}

In Fig.~\ref{fig:1} we show the application of the method to 
the electromagnetic dipole strength in $^{11}$Be. 
The reduced transition probability was deduced from high-energy
${}^{11}$Be Coulomb dissociation at GSI \cite{palit}.
Using a cutoff radius of $R=2.78$~fm and 
an inverse bound-state decay length of
$q=0.1486$~fm${}^{-1}$ as input parameters we extract
an ANC of $C_{0}=0.724(8)$~fm${}^{-1/2}$ 
from the fit to the experimental data. The ANC
can be converted to a spectroscopic factor of $C^{2}S=0.704(15)$
that is consistent with results from other methods.
In the lowest order of 
the effective-range expansion the phase shift 
in the partial wave with orbital angular momentum $l$ and
total angular momentum $j$
is written as $\tan \delta_{l}^{j}= -(x c_{l}^{j}\gamma)^{2l+1}$. 
The halo expansion parameter is denoted by $\gamma=qR=0.4132<1$ 
and $x=k/q
=\sqrt{E/S_{n}}$ where the neutron separation energy 
is given by $S_{n}$. 
The effective range term $\frac{1}{2} r_l k^2$ term can be neglected,
since it leads to a contribution with an extra $\gamma^2$ factor
which is small in the halo nucleus limit $\gamma \to 0$
(at least in the case where the scattering length 
and the effective range parameter are of natural
order).
The dimensionless parameter  $c_{l}^{j}$ corresponds to the scattering
length $a_{l}^{j} = (c_{l}^{j}R)^{2l+1}$. We obtain
$c^{3/2}_{1}=-0.41(86,-20)$ and $c^{1/2}_{1}=2.77(13,-14)$
, corresponding to $a^{1/2}_{1}=456(61,-73)$fm$^3$. This 
unnaturally large value was found
directly from the analysis of  Coulomb dissociation data \cite{palit}. 
It finds a natural explanation in the existence of the $\frac{1}{2}^{-}$
-bound state close to the neutron breakup threshold in ${}^{11}$Be.

 The connection of the scattering length $a_l$ and the bound state
parameter q for $l>0$ is given by
$a_l=\frac{2 (2l-1) R^{2l-1}}{q^2(2l+1)!!(2l-1)!!}$
in a square well model with radius R.
This is a generalization of the
well-known relation $a_0=1/q$ for $l=0$.
The $p_{1/2}$ channel in $^{11}$Be is an example for the 
influence of a halo state on the continuum. 
The binding energy
of this state is given by 184 keV, which corresponds to $q=0.094$~fm$^{-1}$. 
With $R=2.78$~fm one has $\gamma^2=0.068$.
For $l=1$ one has $a_1=\frac{2R^3}{3\gamma^2}=210$~ fm$^3$. 
Applications to low-lying single neutron halo strength
in carbon isotopes are given in \cite{usha}. 

Let us now mention very briefly  the proton case \cite{tyba04}.
Only a numerical approach is possible now.
Again one finds that the ANC is not the only factor which determines the 
S-factor: there is also an influence of the scattering length.
In Fig. 2 the influence of the 
potential depth $V_0$ on the strength distribution is shown.
We refer to \cite{tyba04} for further discussion.

\begin{figure}[tb]
\epsfysize=9.0cm
\begin{center}
\begin{minipage}[t]{8 cm}
\epsfig{file=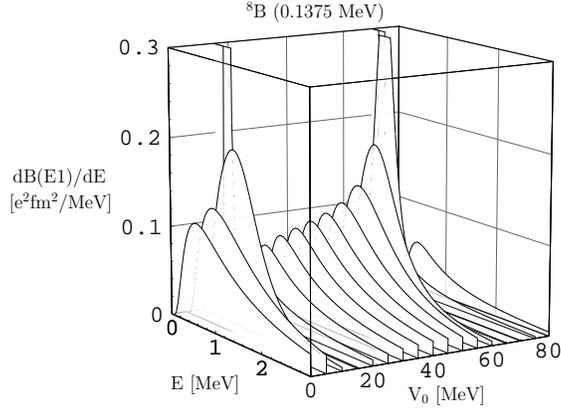,scale=0.5}
\end{minipage}
\begin{minipage}[t]{16.5 cm}
\caption{Reduced transition probablity $dB(E1)/dE$ for the breakup
of $^8$B into a proton and $^7$Be as a function of the 
relative  energy E of the final state for various depths $V_0$
of the potential in the continuum. The proton separation energy 
in $^8$B is given by 0.1375 MeV. For further explanations 
see \cite{tyba04}.
\label{fig2}}
\end{minipage}
\end{center}
\end{figure}

\subsection{Two neutron halo nuclei, application to $^{11}$Li}
While single particle halos show rather simple features
which can largely be explained in the frame of 
the one-body Schr\"odinger equation,
two neutron (or two proton, or neutron-proton) halo nuclei
show many more interesting aspects. 

A Borromean nucleus is a three-body system
which falls apart when one of the particles is removed,
$^{11}$Li is an example. It can be considered as a bound system
of two neutrons and a core ($^9$Li). The system
is barely bound ($E_b\sim 300$keV) and the core can be 
assumed to be inert. The subsystems (n-n) and ( $^9$Li-n)
are unbound. The binding of the Borromean nucleus 
is due to the attractive n-n interaction which leads to
the binding of the three-body system, see e.g. \cite{migdal}.

 Essentially due to the small binding energy the low-lying E1 strength
shows simple universal features as we will show now. 
The system is probed by  
long wavelength photons and can be characterized by a few 
low energy constants. As we have seen in the previous subsection,
the dipole matrixelement is dominated by 
contributions from distances
larger than the range of the core-neutron interaction.
In this region the wave function depends on the binding energy
$E_b$. In a mean field approach the binding energy is shared
equally between the two neutrons. Each 
neutron is bound by $\epsilon=\frac{\hbar ^2 q^2}{2m}=E_b/2$.
The slope of the wave function is characterized by $q$,
the overall magnitude by an asymptotic normalization constant $C_l$:
$\phi_{l,\epsilon}(r) \rightarrow C_l e^{-qr}$.

For simplicity let us neglect spin-orbit forces and assume that
the neutron spin state is the antisymmetric $S=0$ state.
The space part is symmetric , we include $s^2$ and $p^2$
configurations, with the corresponding spectroscopic factors $C^2S_l$.
The radial part of the two-neutron halo wave function is written as
$ \sim b_l \phi_{l,\epsilon} (r_1) 
\phi_{l,\epsilon}(r_2)$.
The amplitudes $b_l$ are related to the spectroscopic factors
by $C^2S_l=|b_l|^2$.
The wave function for $r_{1,2}>R$ is rather simple, it must be
borne in mind that the binding itself is caused by the 
full three-body dynamics. 

The dipole operator is a one-body operator. It is given 
in terms of the Jacobi coordinates $\vec \rho_1$ and 
$\vec \rho_2$ by 
\begin{equation}
M(E1\mu)=Z_{\rm eff}e(\rho_1 Y_{1\mu}(\hat \rho_1)+
\rho_2 Y_{1\mu}(\hat \rho_2))
\end{equation}
The introduction of the Jacobi coordinates introduces
an asymmetry between the two identical neutrons.
This translates into an asymmetry of the effective charges
for both neutrons, we neglect this effect, which is
of the order of $1/A_c$, where $A_c$ is the mass of the core.
For the effective charge we take $Z_{\rm eff}=\frac{-Z_c m_n}{m_n + m_c}$.
(The other possible choice would be $Z_{\rm eff}=\frac{-Z_c m_n}{2 m_n + m_c}$.) 
The final state is given by two continuum neutrons
with momenta $\vec k_1$ and $\vec k_2$. We
assume that they are uncorrelated. Low lying strength
is a long wavelength phenomenon, and the interaction can be
characterized by the n-core scattering lengths $a_0$ and
$a_1$ in the 
$l=0$ and $l=1$ channels. We know that $^{10}$Li is unbound,
so these scattering lengths are negative.
The continuum radial wave functions are given by a 
product of single particle wave functions of the type
given by eq. 3 above. 

The dipole matrix element consists of a product of two terms:
one is characterized by the dipole selection rule
$\Delta l=1$ and is of the type encountered in the
single particle case above. 
We have
\begin{equation}
\frac{dB(E1)}{dE}=Z_{eff}^2\frac{2m}{\pi\hbar^2}
\frac{3}{4\pi}(l_i010|l_f0)^2\frac{C_l^2}{q^5}S_{l_i}^{l_f}
\end{equation}
where the shape 
functions \cite{tybaprl}
are given in terms of the dimensionless quantity $x=\frac{k}{q}$ by
\begin{equation}
S_0^1(\lambda=1)=\frac{4x^3}{(1+x^2)^4}(1-a_1 q^3(1+3x^2)+...)
\end{equation}
for s to p-transitions and by 
\begin{equation}
S_1^0(\lambda=1)=\frac{x(3+x^2)^2}{(1+x^2)^4}(1-\frac{4a_0q}{3+x^2}+...)
\end{equation}
for p-to s-transitions. 

The other term is an overlap term
\begin{equation}
\frac{dN}{dE} = \frac{2}{\pi}
\frac{m}{\hbar^2k} |<\phi_{l\epsilon}|\phi_{l\vec k}>|^2
\end{equation}
where $E=\frac{\hbar^2 k^2}{2m }$ (using the 
normalization conventions of \cite{tyba04}).
One can describe the physics of this matrixelement in quite simple terms:
Due to the sudden removal of one of the neutrons by the 
interaction of the photon with the core, the remaining neutron 
finds itself in the field of the $^9$Li core, where it is no longer bound.
A 'shake off' occurs , which is described by the overlap
term, eq.9. It is non-zero, because the two wave functions
belong to different Hamiltonians. A similar
and familiar example is the 
shake-off of electrons after  $\beta$-decay, where the
nuclear charge is suddenly changed by one unit.
This approach can also be applied to non-Borromean nuclei.
In this case the neutron can also be left in a bound
state with a certain probability which can be 
obtained from the corresponding overlap matrixelement.

The overlap integral is strongly peaked around zero energy. 
For $l=0$ one obtains
\begin{equation}
|<\Phi_{0\epsilon}|\Phi_{0 k}>|^2=\frac{2qk^2(a_0q-1)^2}{(1+(a_0 k)^2)(q^2+k^2)^2}
\end{equation}
In the unitary limit ($a_0 \rightarrow \infty$) this reduces to
\begin{equation}
|<\Phi_{0\epsilon}|\Phi_{0 k}>|^2=\frac{2qk^2}{(q^2+k^2)^2}
\end{equation}
This is probably a good approximation for the overlap term
in the $^{11}$Li case where one neutron is almost bound to
the $^9$Li core.
The analytical formula for
all values of the angular momentum is given in Appendix A.4 of \cite{tyba04}
($\lambda=0$).

Low-lying E1 strength in $^{11}$Li was observed 
experimentally in Ref. \cite{naka}. The 
dipole strength distribution as a function of the relative energy 
$E=E_1+ E_2=\frac{\hbar^2}{2m}(k_1^2+k_2^2)$ 
of the $^9$Li-n-n-system is shown in Fig.3 there. 
It will be interesting to compare this experimental result
to the present approach.
The dipole strength distribution $\frac{dB(E1)_{\rm 2n-halo}}{dE}$ 
depending on the relative 
energy E can be obtained by folding the dipole 
distribution with the overlap term:
\begin{equation}
\frac{dB(E1)_{\rm 2n-halo}}{dE}= 2 \int \int dE_1 dE_2 \frac{dBE1}{dE_2} 
\frac{dN}{dE_1} \delta(E-(E_1 + E_2))
\end{equation}
The factor of 2 arises because there are two neutrons in the same shell.
  We assume that the 
binding energy $E_b$ is known. The shapes of the 
functions $S_0^1$ and $S_1^0$ (see eqs. 7 and 8)
(and $S_1^2$) show characteristic
differences, so the fit to the experimental results
\cite{naka} will be quite sensitive to the spectroscopic factors 
$C^2S_0$ and $C^2S_1$.
The width of the nonorthogonality term depends on the 
scattering lengths, so they can also be determined by this 4-parameter
fit, but probably one will be less sensitive to these quantities.
The experimental results \cite{naka} are quite well explained by the theory of 
Esbensen and Bertsch \cite{bees}. It will be interesting to compare the present
approach also to the more detailed theory of \cite{bees}. 
In the present approach only a few low energy parameters like the 
scattering lengths in the relevant channels enter.
  
\section{Trojan Horse Method}

A similarity between cross sections for two-body and closely
related three-body reactions under certain kinematical conditions
\cite{Fuc71}
led to the introduction of the Trojan-Horse method 
\cite{Bau84,Bau76,Typ00,tyba02}.
In this indirect approach a two-body reaction
\begin{equation} \label{APreac}
 A + x \to C + c
\end{equation}
that is relevant to nuclear astrophysics is replaced by a reaction
\begin{equation} \label{THreac}
 A + a \to C + c + b
\end{equation}
with three particles in the final state. 
One assumes that the Trojan horse
$a$ is composed predominantly of clusters $x$ and $b$, i.e.\  $a=(x+b)$. 
This reaction can be considered as a special case of a transfer 
reaction to the continuum. It is studied experimentally under quasi-free
scattering conditions, i.e.\ when the momentum transfer to the
spectator $b$ is small. The method was primarily applied to the
extraction of the low-energy cross section of reaction
(\ref{APreac}) that is relevant for astrophysics. However, the method
can also be applied to the study of single-particle states in exotic
nuclei around the particle threshold.

\subsection{Continuous Transition from positive
to negative energies in the (A+x)-channel} 
Motivated by this 
we look again at the relation between transfer to 
bound and unbound states. We study the reaction
$A+a \rightarrow B+b$
where $a=(b+x)$ and B denotes the final
$B=(A+x)$ system. It can be a bound state
$B$ with binding energy $E_{\rm bind}=-E_{Ax}(>0)$,
the open channel $A+x, \mbox{with }E_{Ax}>0$, or 
another channel $C+c$ of the system $B=(A+x)$.
In particular, the reaction $x+A \rightarrow C+c$ can have
a positive $Q$ value and the energy $E_{Ax}$ can be negative
as well as positive.
As an example we quote the recently studied Trojan horse reaction
$d$+${}^{6}$Li \cite{auro03} applied to the ${}^{6}$Li$(p,\alpha)^{3}$He
two-body reaction (the neutron being the spectator).
In this case there are two charged
particles in the initial state (${}^{6}$Li+$p$).
Another example with a neutral particle $x$
would be ${}^{10}$Be$ + d \rightarrow p
+{}^{11}$Be$ +\gamma$.  
The general question which we want to answer
here is how the two regions $E_{Ax}>0$ and
$E_{Ax}<0$ are related to each other. 
E.g., in Fig.\ 7 of \cite{auro03} the coincidence yield 
is plotted as a function of the ${}^{6}$Li-$p$ relative energy.
It is nonzero at zero relative energy. How does 
the theory \cite{tyba02} (and the experiment)
continue to negative relative energies?
With this method, subthreshold resonances can be 
investigated rather directly.


 The Trojan Horse amplitude f involves the combination
$f \propto S_{lc} \cdot J_l^+$ (see eq. (61) of \cite{tyba02}).
In this reference, only the case  $E \equiv E_{Ax} >0$ is dealt 
with explicitly.
The threshold behaviour of $J_l^+$ in the case of neutrons is given by (A.30),
for charged particles by eq. (59) of \cite{tyba02}, respectively.
$S_{lc}$ is the S-matrix-element connecting the channel $c \equiv c+C$
and the elastic channel $A+x$ with angular momentum $l$. 
For neutral particles 
the threshold behaviour is $|S_{lc}|^2 \sim (kR)^{2l+1}$ and the cross
section $\sigma \propto k^{-3} |S_{lc}J_l^+|^2$ tends to a finite constant.
A similar analysis can be done for charged particles. 
Let us now study the case $E_{Ax}<0$ and the transition from $E_{Ax}>0$ 
to $E_{Ax}<0$. 
In the latter case the wave number k becomes complex, k$=i$q and
the factor $J_l^+$ involves a decaying function,
instead of the oscillatory one. The quantity $S_{lc}$ is no longer an 
S-matrixelement, it is the normalization factor for the decaying
wave-function.
The continuous transition for both cases was studied in \cite{sanser}
in a two-channel model. 

An 
interesting case is when there are subthreshold resonances,
like in the systems 
$^{14}N(p,\gamma)^{15}O$ and in $^{12}C(\alpha,\gamma)^{16}O$.
In \cite{msuproc,sanser} a two-channel model is studied.The standard Breit-Wigner
result is obtained
\begin{equation}
 S_{ij}=e^{i( \xi_i +\xi_f)} (\delta_{ij}-\frac{i\sqrt{\Gamma_i \Gamma_j}}
{E-E_R+ i \Gamma /2})
\end{equation}
where $E_R$ denotes the resonance energy, the partial widths
are given by $\Gamma_1$ and $\Gamma_2$, the total width is given
by $\Gamma=\Gamma_1 + \Gamma_2$.

For $E<0$ the width $\Gamma_1$
turns out to be imaginary \cite{sanser}. 
For $E_R>0$ we have a resonance, for $E_R<0$ a 
subthreshold resonance, the formulae are valid for both cases.
In \cite{brune1, brune2} sub-Coulomb $\alpha$ transfer is used to study
the low energy $^{12}C(\alpha,\gamma)^{16}O$ S-factor. Transitions to the 
$1^-$ and $2^+$ bound states are observed and the reduced $\alpha$ widths 
of these states is obtained. 

\section{Conclusion}
While the foundations of direct reaction theory
have been laid several decades ago, the new possibilites which have 
opened up with the rare isotope beams 
are an invitation to revisit this field. The general frame 
is set by nonrelativistic many-body quantum scattering theory,
however, the increasing level of precision demands
a good understanding of relativistic effects notably in 
intermediate-energy Coulomb excitation.

The properties of halo nuclei 
depend very sensitively on the binding energy 
and despite  the ever increasing
precision of microscopic approaches using
realistic NN forces it will not be possible, say,
to predict the binding energies of nuclei to a level of 
about 100~keV.
Thus halo nuclei ask for 
new approaches in terms of some effective 
low-energy constants.
Such a treatment was provided in Ch.\ 2 and an example
with the one-neutron halo nucleus $^{11}$Be was given.
With the future radioactive beam facilities at RIKEN and
GSI one will be able to study also neutron halo
nuclei for intermediate masses in the years to come. 
This is expected to
be relevant also for the astrophysical r-process.
It is a great challenge to extend the present approach
for one-nucleon halo nuclei to more complicated cases,
like two-neutron halo nuclei. In the present work we
obtained a rather simple formula for two-neutron halo
nuclei with a transparent interpretation: after the first neutron
is removed, the second one is left in a different
field (Hamiltonian) which leads to a shake-off process,
similar to the rearrangement of the atomic cloud after an
$\alpha$- or $\beta$ decay. 

While the Coulomb dissociation method relies essentially
only on QED, precise experiments can give,
in combination with a thorough theoretical analysis,
precise answers for the astrophysical S factors. 
In the Trojan-Horse Method
more phenomenological aspects enter, 
like optical model parameters and effective nuclear interactions.
This makes the interpretation of the experimental
results in terms of astrophysical S factors 
less precise. In the Trojan-Horse Method one can extract
bare nucleus (unscreened) S-factors, see e.g. \cite{auro03}.  
In the interpretation of screening effects one relies on the accuracy of the 
energy dependence. This is certainly better fulfilled than 
the accuracy of the
absolute values.



\bibliographystyle{aipproc}   


%

\end{document}